\documentclass[%
 twocolumn,
 amsmath, amssymb,
 aps,
 prl,
]{revtex4-1}

\usepackage{graphicx}
\usepackage{dcolumn}
\usepackage{bm}
\usepackage[colorlinks]{hyperref}

\usepackage{multirow}
\usepackage{diagbox}
\renewcommand{\vec}[1]{{\bf #1}}
\newcommand{\braket}[1]{\langle #1  \rangle}
\newcommand{\ket}[1]{| #1  \rangle}

\newcommand{\qbraket}[1]{[#1]}

\newcommand{\site}[1]{{[#1]}}
\newcommand{\tmp}{\text{tmp}}
\newcommand{\Tr}{\text{Tr}}

\renewcommand{\min}{\text{min}}
\newcommand{\Potts}{\text{Potts}}
\newcommand{\Log}{\text{Log}}
\newcommand{\abs}[1]{|#1|}

\begin{document}

\title{Dynamical Quantum Phase Transition of the Quantum $N$-state Potts Chain with Quenched Disorder}

\author{Yantao Wu}

\affiliation{
The Department of Physics, Princeton University
}

\date{\today}
\begin{abstract}
We present an exact renormalization group analysis of the Loschmidt amplitude of the quantum $N$-state Potts chain with random quench-disordered nearest neighbor bonds, under the extreme dynamical quantum quench.
We prove that the phase transition of the Loschmidt rate function remains sharp in general.
For typical bond distributions, the phase transition is found to be a linear-cusp, as in the pure model.  
For some special discrete bond distributions, however, the rate function exhibits logarithmic divergences. 
These singularities are due to the competition between the non-critical dynamical phases of the pure model, which is very different from how disorder affects equilibrium phase transitions.  
In addition, due to the periodicity of the complex exponential function, all continuous bond distributions result in rate functions which converge to a universal value at large time. 
\end{abstract}

\pacs{Valid PACS appear here}
\maketitle
Due to the rapid progress in experimental techniques, dynamical quantum phase transitions (DQPT) have received lots of interest recently \cite{Heyl_2018}. 
It was first found in \cite{dqpt} that the quantum dynamics of the transverse field Ising chain can exhibit singular dependence on time in the thermodynamic limit.   
Subsequently the DQPT has been studied in many other examples \cite{Heyl_2018, zvyagin}. 
The central quantity in DQPT that hosts this singular behavior is called the Loschmidt amplitude $G(t)$:   
\begin{equation}
  G(t) = \braket{\psi_0|e^{-iHt}|\psi_0}
\label{eq:Loschmidt}
\end{equation}
where $\ket{\psi_0}$ is a quantum state evolving under the Hamiltonian $H$ for time $t$.  
When $\ket{\psi_0}$ is not an eigenstate of $H$, $G(t)$ measures the return probability of the system due to a sudden change in the system Hamiltonian. 
Due to the lack of organizing principles, such as the minimization principle of the free energy, and numerical tools, such as the Monte Carlo simulation, DQPTs present many challenging problems.    
One such problem concerns with how disorder inherent in the Hamiltonian affects the DQPT of the pure model.  
While the study of quench-disordered systems has a rich tradition in equilibrium phase transitions \cite{cardy}, the systems studied for DQPTs are almost all spatially homogeneous, perhaps mostly due to the young age of DQPTs.  
It is currently not clear how disorder affects DQPTs on a general level \cite{Heyl_2018}, and the analytical knowledge on any non-trivial example would be desirable. 

In this paper, we initiate the quantum dynamics with the ground state of the Potts Hamiltonian with infinite transverse-field, and then evolve it under the Hamiltonian with zero transverse-field.  
The effect of disorder on this DQPT is studied. 
We deal with systems whose nearest-neighbor bonds are drawn independently from a probability distribution.     
Relying on the knowledge of the non-equilibrium renormalization group (RG) fixed points of the pure Potts chain obtained recently \cite{potts_rg}, we will prove that in general the DQPT in the presence of random disorder remains sharp. 
For most distributions of the random bonds, the DQPT will be a linear-cusp, as in the pure model \cite{scaling_tfic, potts_rg}.
For some fine-tuned discrete distributions, however, the DQPT exhibits logarithmic divergences. 
In addition, the rate function reaches a universal plateau value at large time for all continuous bond distributions. 

Consider the $N$-state Potts chain of $L$ sites with periodic boundary condition with the Hamiltonian \cite{dqpt_potts}, 
\begin{equation}
  H_{\Potts} = -\sum_{i=1}^L J_i(\sigma_i^\dag \sigma_{i+1} + \sigma_{i+1}^\dag \sigma_i) - f\sum_{i=1}^L(\tau_i^\dag + \tau_{i})
\end{equation}
where the operators $\sigma_i$ and $\tau_i$ act on the $N$ states of the local Hilbert space at site $i$, which we label by $\ket{0}_i,...,\ket{m}_i,...\ket{N-1}_i$.  
In this local basis, the $\sigma_i$ is a diagonal matrix with diagonal elements $\omega^m$ where $\omega = e^{i2\pi/N}$ and $m = 0, \cdots, N-1$. 
$\tau_i$ permutes $\ket{0}_i\rightarrow \ket{1}_i, \ket{1}_i \rightarrow \ket{2}_i$, etc., and together with its adjoint operator acts as a transverse-field.   
The bond strength $J_i$ will be different at different lattice sites. 
For the Loschmidt amplitude, we take the paramagnetic direct product state $\ket{\psi_0} = \otimes_{i=1}^L \frac{1}{\sqrt{N}}(\ket{0}_i + \ket{1}_i + ... + \ket{N-1}_i)
$ and the ferromagnetic Hamiltonian $H = -\sum_{i}J_i (\sigma_i^\dag \sigma_{i+1} + \sigma_{i+1}^\dag \sigma_i)$. 
In this case, $G(t)$ becomes formally identical to a classical partition function \cite{scaling_tfic, potts_rg}:  
\begin{equation}
\begin{split}
  G(t) &= \frac{1}{N^L} \sum_{\vec m} ... T_{m_im_{i+1}}^{\site{i}} T_{m_{i+1}m_{i+2}}^{\site{i+1}} T_{m_{i+2}m_{i+3}}^{\site{i+2}}... 
\end{split}
\end{equation}
where $\vec m = \{m_1, m_2, ..., m_{L}\}$ is the set of degrees of freedom of this partition function and $m_i = 0, 1, ..., N-1$ takes the value of a Potts spin at site $i$.   
Here $T^\site{i}_{m_i m_{i+1}}$ is the transfer matrix of the system between sites $i$ and $i+1$ and depends only on the difference between $m_i$ and $m_{i+1}$ modular $N$, $m \equiv (m_{i+1} - m_{i}) | N$ \cite{potts_rg}. 
That is, 
\begin{equation}
  T^\site{i}_{m_i m_{i+1}} \equiv E_{m} = e^{it J_i 2 \cos(\frac{2\pi}{N}m)} 
  \label{eq:transfer}
\end{equation}
The scaling of $G(t)$ with $L$ is such that the following rate function is intensive in the thermodynamic limit \cite{dqpt}:
\begin{equation}  
  l(t) = -\frac{1}{L} \log \abs{G(t)}^2 = -\frac{2}{L} \Re\{\Log \, G(t)\}
\end{equation}
where $\Log$ is the principal complex logarithmic function. 
The rate function is thus the analog of free energy per site in DQPTs. 
To analyze $l(t)$, we perform the decimation coarse-graining on the Potts chain, i.e. every other spin is decimated away while keeping $G(t)$ invariant.   
This coarse-graining procedure is equivalent to multiplying two neighboring transfer matrices into one.  
In \cite{potts_rg}, it has been explained that the natural coupling constants to perform the RG procedure are the $E_m$s. 
As a result of matrix multiplication of two neighboring transfer matrices, the coupling constants are renormalized as \cite{potts_rg}:   
\begin{equation}
  \begin{split}
    \text{step 1: }& E'^{\site{i'}}_{m,\text{tmp}} = \sum_{l = 0}^{N-1} E^\site{i}_l E^\site{i+1}_{m-l}
  \\
  \text{step 2: }& E'^{\site{i'}}_{m} = \frac{E'^{\site{i'}}_{m,\text{tmp}}}{E'^{\site{i'}}_{s, \text{tmp}}} 
\end{split}
\label{eq:RG}
\end{equation}
where $E'^{\site{i'}}_{s, \text{tmp}}$ is the first nonzero $E'^{\site{i'}}_{m,\text{tmp}}$, counting $m$ from $0,1,..$ to $N-1$. 
Step 2 of Eq. \ref{eq:RG} is necessary for the existence of an RG fixed point, and never becomes singular for the homogeneous Potts chain, which, as we will see, will not be true for a disordered chain. 

Consider the chain where the nearest-neighbor bonds at different lattice sites are drawn independently from a probability distribution, $P(J_i)$.  
Since it is the free-energy that self-averages in the equilibrium statistical mechanics of random systems \cite{cardy}, we expect that the self-averaging quantity that one should study here is the quench-averaged rate function:  
\begin{equation}
  \qbraket{l(t)} = \int d\vec J P(\vec J) l_{\vec J}(t)
\end{equation}
where $\vec J = \{J_1, J_2, ...\}$ is one realization of the bonds with a rate function $l_{\vec J}(t)$, and $P(\vec J) = \Pi_i P(J_i)$ is the probability density of this realization. 
$\qbraket{\cdot}$ denotes the quench-averaging under $P(\vec J)$. 
The loss of translational invariance and the fact that one has to take $\Log$ before doing the quench-averaging bring significant challenge to the computation of $[l(t)]$.  

For equilibrium questions, RG is a powerful tool to analyze phase transitions of disordered systems \cite{lubensky, cardy}.    
In our case, however, the RG equation (Eq. \ref{eq:RG}) has a fatal problem: it becomes singular when renormalizing the two stable fixed-points of the pure system.   

For concreteness, let us take $N = 2$ and generalize the results later for other $N$s. 
When $N = 2$, the coupling constants can be made all real by coarse-graining the transfer matrix once: 
\begin{equation}
  \begin{split}
    T^\site{i'} &= \begin{pmatrix} e^{i2J_it} & e^{-2J_it} \\ e^{-i2J_it} & e^{2J_it} \end{pmatrix}  \begin{pmatrix} e^{i2J_{i+1}t} & e^{-2J_{i+1}t} \\ e^{-i2J_{i+1}t} & e^{2J_{i+1}t} \end{pmatrix} \\ 
    &= 2 \begin{pmatrix} \cos(2(J_i + J_{i+1})t) & \cos(2(J_i - J_{i+1})t) \\ \cos(2(J_i - J_{i+1})t) & \cos(2(J_i + J_{i+1})t)\end{pmatrix}.
 \end{split}
 \label{eq:real}
\end{equation}
There are two stable non-equilibrium RG fixed points, $\vec E^*_a = (1, 1)$ and $\vec E^*_b = (1, -1)$, for the pure model \cite{potts_rg}. 
The attractive basin for $\vec E^*_a$ is $\vec E = (1, a), a > 0$, and for $\vec E^*_b$ is $\vec E = (1, b), b < 0$. 
After step 1 of Eq. \ref{eq:RG} of the coupling constants at two lattice sites, $\vec E^\site{i} = (1, x_i)$ and $\vec E^\site{i+1} = (1, x_{i+1})$, one obtains $\vec E'^{\site{i'}}_\tmp = (1+x_ix_{i+1}, x_i + x_{i+1})$.    
Thus, within the attractive basin of each non-critical fixed-point, $E'^{\site{i'}}_{0,\tmp} \ge 1$ and the RG equation is perfectly regular.  
In addition, as long as both of $\vec E^\site{i}$ and $\vec E^\site{i+1}$ are in the same attractive basin, their renormalized coupling constant will be closer to the respective stable fixed-point than either $\vec E^\site{i}$ or $\vec E^\site{i+1}$. 
However, when $\vec E^\site{i} = \vec E^*_a$ and $\vec E^\site{i+1} = \vec E^*_b$, step 1 of Eq. \ref{eq:RG} gives, in the form of transfer matrices,  
\begin{equation}
  \begin{pmatrix}1 & 1 \\ 1 & 1\end{pmatrix}
  \begin{pmatrix}1 & -1 \\ -1 & 1\end{pmatrix} 
= 
  \begin{pmatrix}0 & 0 \\ 0 & 0\end{pmatrix}, 
\end{equation}
which makes the second step of Eq. \ref{eq:RG} singular. 
As the RG procedure proceeds, the coupling constants of the disordered chain very quickly settle into the vicinity of one of the two stable fixed-points, and the RG procedure eventually fails.   

To overcome this failure, one first notes that the normalized transfer matrices at different sites commute:   
\begin{equation}
  \begin{pmatrix}1 & x_i \\ x_i & 1\end{pmatrix}
    \begin{pmatrix}1 & x_{i+1} \\ x_{i+1} & 1\end{pmatrix} 
= 
\begin{pmatrix}1 & x_{i+1} \\ x_{i+1} & 1\end{pmatrix}
  \begin{pmatrix}1 & x_{i} \\ x_{i} & 1\end{pmatrix}. 
\end{equation}
Consequently, we can move all the transfer matrices in the phase of $\vec E^*_a$ to the left side of the chain, and those in the phase of $\vec E^*_b$ to the right side without changing the value of $l_{\vec J}(t)$.  
The $\vec E^*_a$ and $\vec E^*_b$ side of the chain can then be respectively renormalized into one transfer matrix without incurring any singularity:   
\begin{equation}
  T_a = \begin{pmatrix}1 & 1+\epsilon_a \\ 1 + \epsilon_a & 1\end{pmatrix},   T_b = \begin{pmatrix}1 & -1+\epsilon_b \\ -1 + \epsilon_b & 1\end{pmatrix}
    \label{eq:Ta}
\end{equation}
where if there are sufficiently many transfer matrices on both sides before the renormalization, $\abs{\epsilon_a} \ll 1$ and $\abs{\epsilon_b} \ll 1$.   
In the process, a regular part of the rate function will be extracted due to step 2 of Eq. \ref{eq:RG}. 
All of the singularity of the rate function resides in $T_a$ and $T_b$.  

To clarify the above RG procedure, we decompose the quench-averaged rate function as follows   
\begin{equation}
  \begin{split}
    \qbraket{l(t)} & = l_0 + \qbraket{l_l(t)}  + \qbraket{l_r(t)} + \qbraket{l_s(t)}   
  \end{split}
  \label{eq:decompose}
\end{equation}
where $l_0 = -\frac{2}{L}\log(N^L)$, and $\qbraket{l_l(t)}$ and $\qbraket{l_r(t)}$ are the two regular parts extracted from $\qbraket{l(t)}$ by the RG procedure on the two sides of the chain.   
$\qbraket{l_s(t)}$ is the singular part of the rate function and is given by  
\begin{equation}
\begin{split}
  \qbraket{l_s(t)} &= -\frac{2}{L}\qbraket{\Re\{\Log\, \Tr(T_a(t)T_b(t))\}} 
\\ 
&= -\frac{2}{L} \qbraket{\log \abs{N(\epsilon_a - \epsilon_b + \epsilon_a\epsilon_b)}} 
\end{split}
\end{equation}
Any chain can also be viewed as an assembly of $n$ chains of length $L_0 = \frac{L}{n}$.   
One can independently renormalize these $n$ parts and will end up with a chain composed of transfer matrices $T_{a,1}, T_{a,2}, \cdots, T_{a,n}$, and $T_{b,1}, T_{b,2}, \cdots, T_{b,n}$.  
These transfer matrices may be different due to the fluctuation in the realization, but are the same in distribution.  
The final $\epsilon_a$ of the full chain will then be 
\begin{equation}
\begin{split}
  \epsilon_a &= \frac{\text{the off-diagonal element of }(T_{a,1}...T_{a,n})}{\text{the diagonal element of }(T_{a,1}...T_{a,n})} - 1  
  \\
  &= \epsilon_{a,1}...\epsilon_{a,n} + \text{higher-order terms} 
\end{split}, 
\label{eq:epsilon_a}
\end{equation}
where $\epsilon_{a,1}$, etc., is defined by  
\begin{equation}
  T_{a,1} \equiv \begin{pmatrix}1 & 1+\epsilon_{a,1} \\ 1+\epsilon_{a,1} & 1\end{pmatrix}.  
\end{equation}
A similar expression can be written for $\epsilon_b$. 
In the thermodynamic limit $L_0 \rightarrow \infty, n \rightarrow \infty, L \rightarrow \infty$, $\epsilon_a$ and $\epsilon_b$ both approach zero, and the singular part of the quench-averaged rate function will be    
\begin{equation}
  \begin{split}
    \qbraket{l_s(t)} &= - \lim_{L \rightarrow \infty} \frac{2}{L} \qbraket{\log(\max(\abs{\epsilon_{a,1}...\epsilon_{a,n}},\abs{ \epsilon_{b,1}...\epsilon_{b,n}}))}  
  \\
  &= - \lim_{L_0 \rightarrow \infty} \frac{2}{L_0} \qbraket{\log(\max(\abs{\epsilon_{a,1}}, \abs{\epsilon_{b,1}}))}  
  \\
  &= \lim_{L \rightarrow \infty}  \qbraket{\min(-\frac{2}{L}\log\abs{\epsilon_{a}}, -\frac{2}{L}\log \abs{\epsilon_{b}})} 
\end{split}
\end{equation}
Here we have used the fact that there is no difference between $\epsilon_a$ and $\epsilon_{a,1}$ in the thermodynamic limit.  
As $\epsilon_a$ and $\epsilon_b$ scale exponentially with $L$, as seen from Eq. \ref{eq:epsilon_a}, the above limit exists, and $\qbraket{l_s(t)}$ can finally be written as    
\begin{equation}
  \qbraket{l_s(t)} = \min(l_{a}(t), l_{b}(t)) 
  \label{eq:l_s}
\end{equation}
where 
\begin{equation}
  l_{a/b}(t) = -\lim_{L\rightarrow \infty}\frac{2}{L} \qbraket{\log\abs{\epsilon_{a/b}(t)}}. 
  \label{eq:la}
\end{equation}
In Eq. \ref{eq:l_s}, the order of $\min$ and $\qbraket{\cdot}$ can be swapped, because of the self-averaging property of $l_a(t)$ and $l_b(t)$.  
Now, here is the point: because $l_a(t)$ and $l_b(t)$ are respectively calculated from the renormalization of the system in the same non-critical phase, they should be smooth functions of $t$ if either one of $\epsilon_a$ or $\epsilon_b$ is non-zero.   
$\qbraket{l_s(t)}$ will thus generically have a linear singularity when $l_a(t)$ and $l_b(t)$ intersect.  
However, when $\epsilon_a$ and $\epsilon_b$ both become zero, the rate function diverges logarithmically. 

\begin{figure}[hth]
\centering
  \includegraphics[scale=0.4]{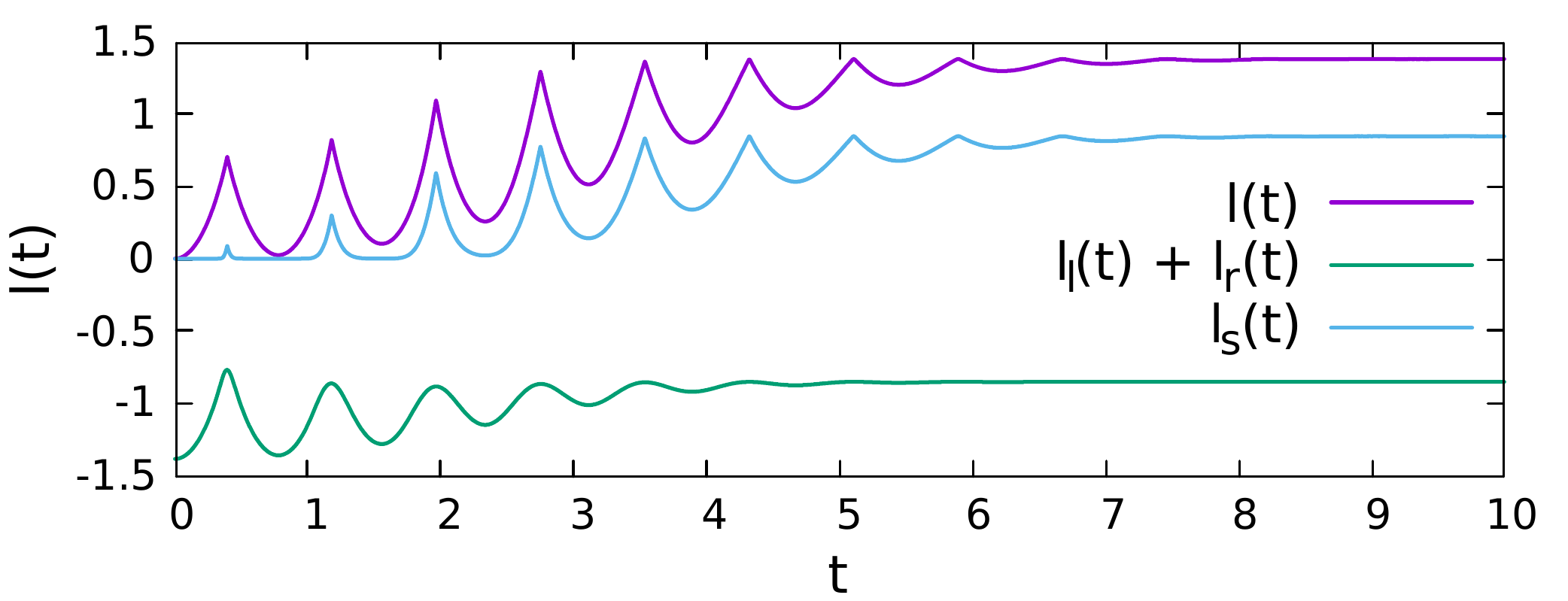}
  \caption{The quench-averaged rate function of the disordered Potts chain defined by Eq. \ref{eq:Ji}.
The calculation is done for $L = 2^{16}$, and is averaged over $2^{10}$ realizations.}
\label{fig:rate_L16}
\end{figure}

\begin{figure}[hth]
\centering
\begin{minipage}{.23\textwidth}
  \includegraphics[scale=0.27]{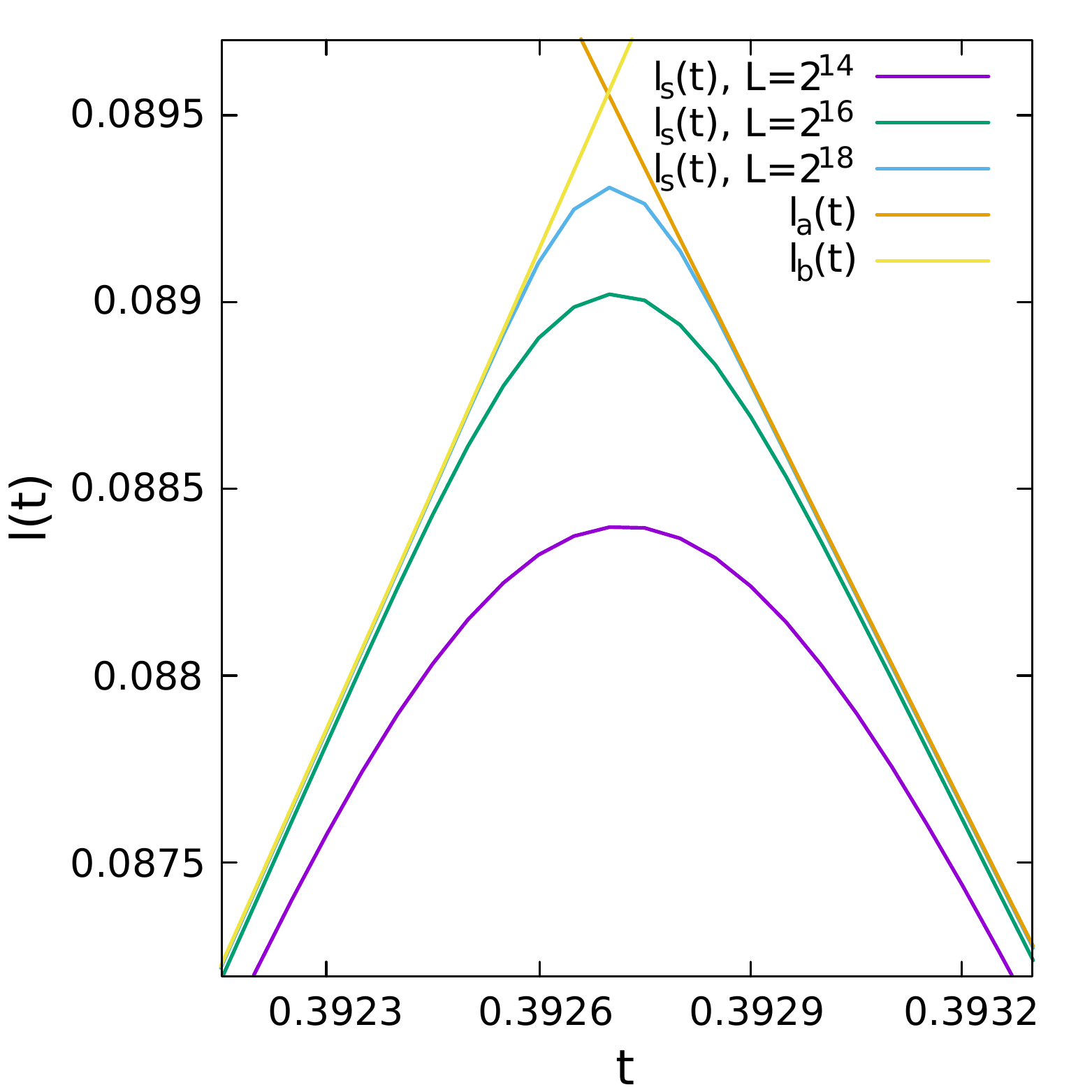}
\end{minipage}%
\begin{minipage}{.23\textwidth}
  \includegraphics[scale=0.27]{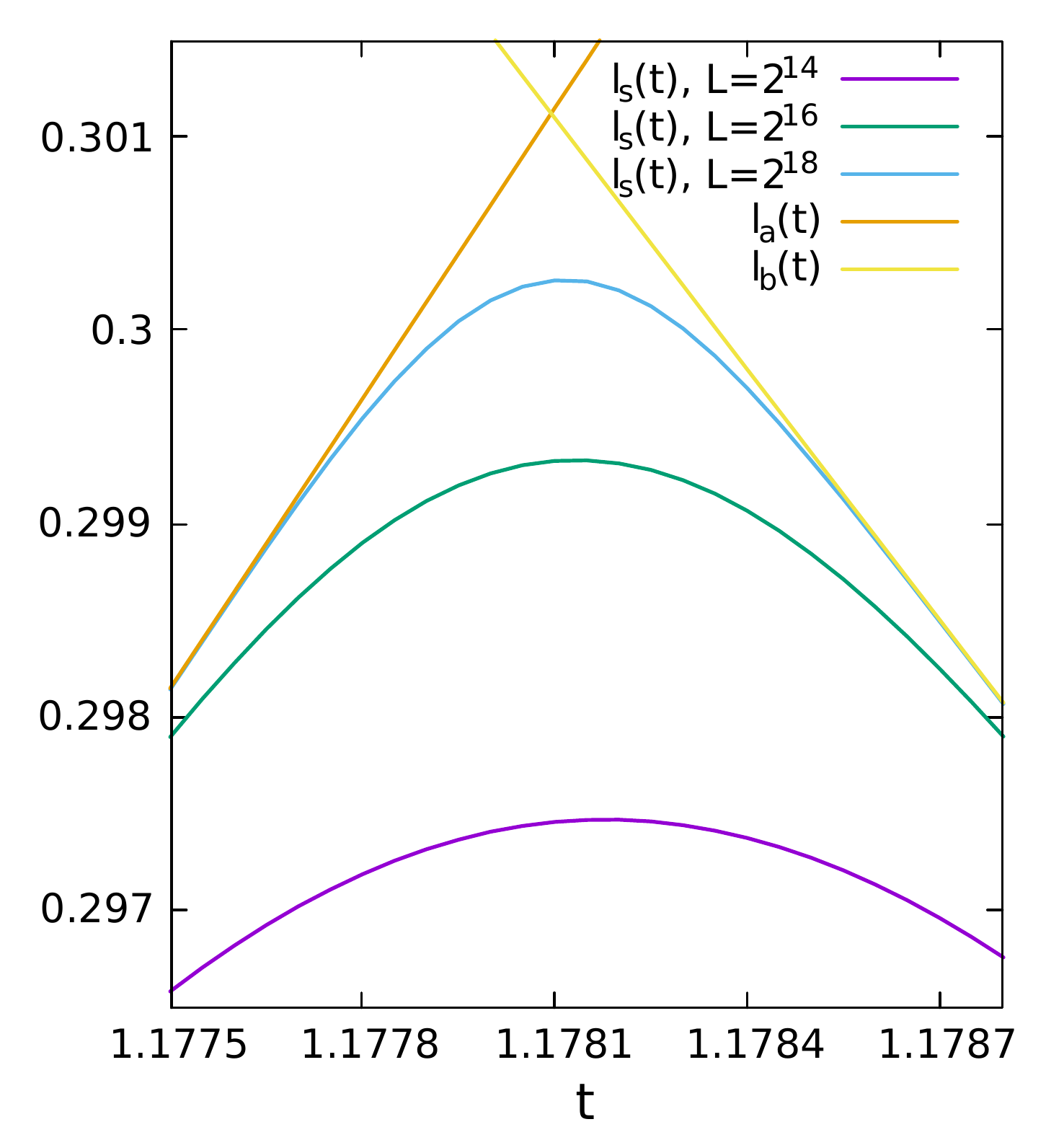}
\end{minipage}%
\caption{$\qbraket{l_s(t)}$, $\qbraket{l_a(t)}$, and $\qbraket{l_b(t)}$ around the first (left panel) and the second (right panel) peaks of the $\qbraket{l(t)}$ in Fig. \ref{fig:rate_L16}.
The $\qbraket{l_s(t)}$ is computed for $L = 2^{14}, 2^{16}$, and $2^{18}$ respectively with $2^{15}, 2^{14}$, and $2^{13}$ realizations.  
The $\qbraket{l_a(t)}$ and $\qbraket{l_b(t)}$ have very weak size dependences and are only shown here for $L = 2^{18}$. 
}
\label{fig:rate_eps}
\end{figure}

Let us first consider an example exhibiting linear singularity, where the random bonds are given by  
\begin{equation}
  J_i = J_0 + J_1 g, \hspace{5mm} g \sim \mathcal{N}(0, 1) 
  \label{eq:Ji}
\end{equation}
independently at each site $i$.
Here $g$ is a unit Gaussian random variable, and $J_0 = 1$ and $J_1 = 0.1$. 
For any realization of the bonds, the various terms of the rate function in Eq. \ref{eq:decompose} can be numerically calculated by the RG procedure.  
An arbitrary precision arithmetic package, such as TTMath \cite{ttmath}, which we use, will be necessary for the calculation of a long chain. 
The result of the calculation is presented in Fig. \ref{fig:rate_L16} and Fig. \ref{fig:rate_eps}.  

\begin{figure}[hth]
\centering
\begin{minipage}{.23\textwidth}
  \includegraphics[scale=0.27]{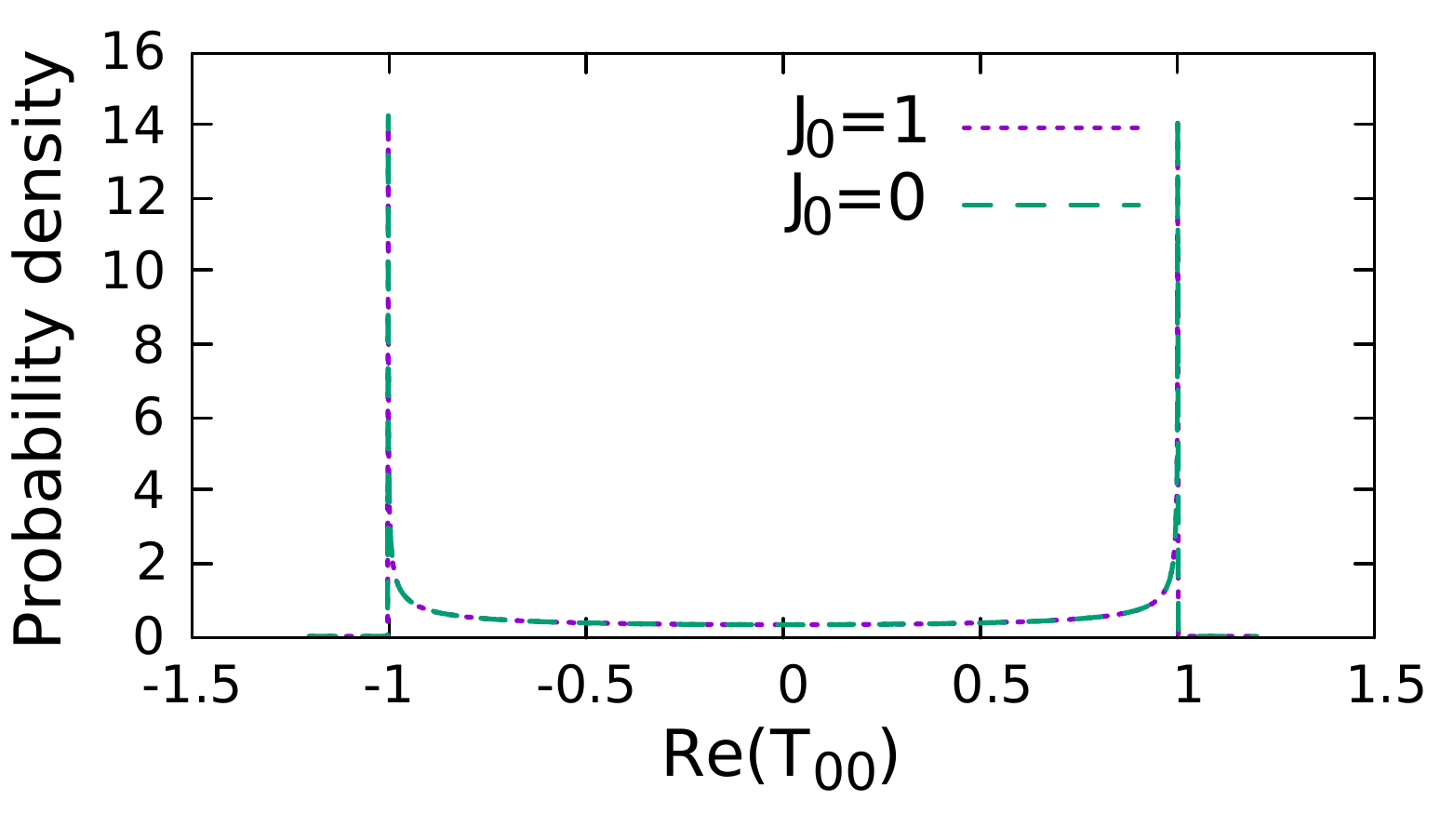}
\end{minipage}%
\begin{minipage}{.23\textwidth}
  \includegraphics[scale=0.27]{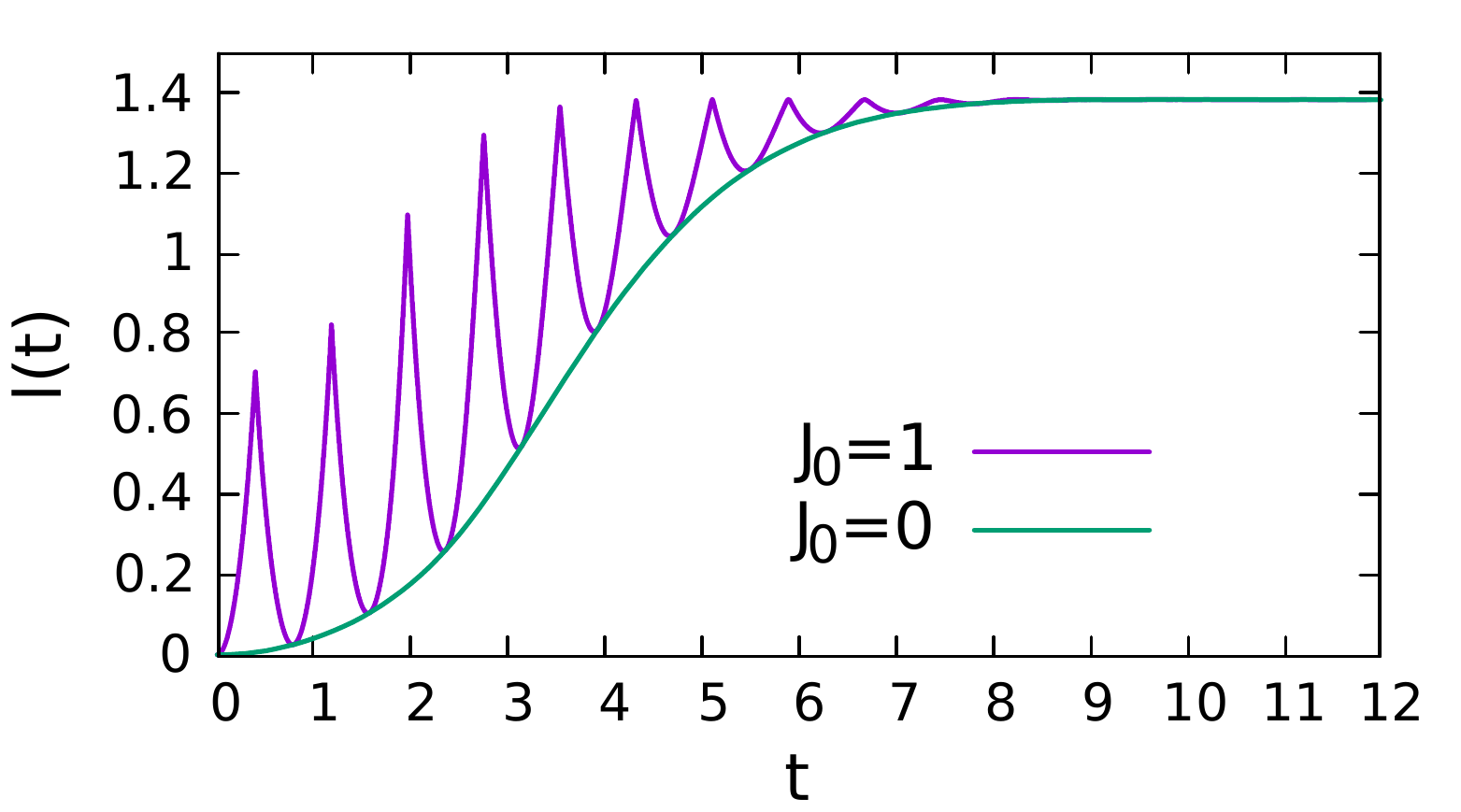}
\end{minipage}%
\caption{Left: The probability density of the $\Re\{T_{00}\}$ for the two disordered chains described by Eq. \ref{eq:J_i} with $J_1 = 0.1$ and $J_0 = 1$ and 0, at $t = 12$. 
Right: The rate function for the same two chains with $L = 2^{16}$. 
}
\label{fig:compare}
\end{figure}

Another feature seen in Fig. \ref{fig:rate_L16} is that the rate function approaches a plateau at large $t$.   
This, in fact, is a general feature in all disordered Potts chain with continuous bond distributions.   
For example, consider the $T^\site{i}_{00}$ element of the transfer matrix in Eq. \ref{eq:transfer} for $N = 2$: $T^\site{i}_{00} = e^{i2tJ_i}$.   
If the distribution of $J_i$ is continuous, then at large $t$ the probability density of $2tJ_i$ within one period of the complex exponential function, i.e. $2\pi$, can be viewed as constant.   
Thus, at large $t$, the distribution of $e^{i2tJ_i}$ approaches the distribution of $e^{i2x}$ where $x$ is distributed uniformly. 
This means not only that the rate function of a disordered chain with a given bond distribution goes into an asymptotic value at large $t$, but also that this asymptotic value is the same for chains with different bond distributions, as long as the bond distribution is continuous.     
For example, in Fig. \ref{fig:compare}, we show the bond distribution and the rate function of the two chains described by Eq. \ref{eq:J_i} with $J_1 = 0.1$ and $J_0 = 1$ and $0$.  

\begin{figure}[hth]
\centering
  \includegraphics[scale=0.4]{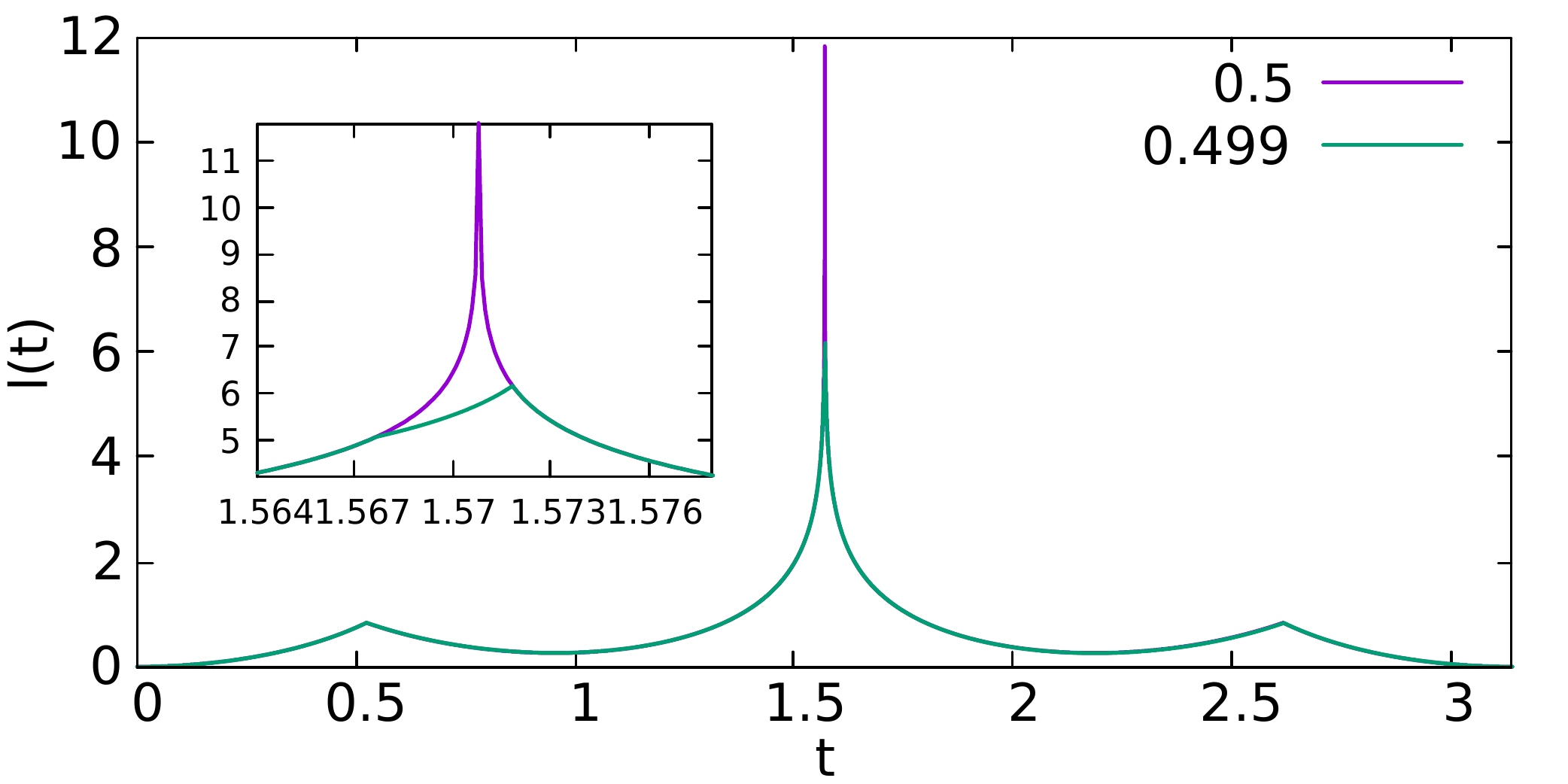}
  \caption{The rate function of a chain with $J_i = 1$ and $0.5\, (0.499)$ each with probability $\frac{1}{2}$.
  Here, because of the commutativity of the transfer matrices, the chain can be made homogeneous by alternatingly putting $1$ and $0.5\, (0.499)$ one after another along the chain. 
  The calculation can thus be done for $L = \infty$ by computing the leading eigenvalue of the transfer matrix of the homogeneous chain. 
  } 
\label{fig:rate_log}
\end{figure}

As one can imagine, almost all distributions of nearest neighbor bonds result in nonzero $\epsilon_a(t)$ and $\epsilon_b(t)$. 
However, if the chain is composed of transfer matrices $T_l$ and $T_r$, which renormalize into $\vec E_a^*$ and $\vec E_b^*$ in finite RG iterations, then both $\epsilon_a(t)$ and $\epsilon_b(t)$ become zero.  
For example, when $N = 2$, consider a chain with nearest neighbor bonds,   
\begin{equation}
  J_i = \begin{cases}1 , &\text{ with probability $p$} \\ \frac{1}{2},  &\text{ with probability $1-p$} \end{cases}
  \label{eq:J_i} 
\end{equation}
where $0 < p < 1$. 
At $t = \frac{\pi}{2}$, $J_i = 1$ and $\frac{1}{2}$ respectively give transfer matrices $T_l$ and $T_r$:   
\begin{equation}
  T_l = \begin{pmatrix}-1 & -1 \\ -1 & -1\end{pmatrix}, \hspace{5mm} T_r = \begin{pmatrix}i & -i \\ -i & i\end{pmatrix} 
\end{equation}
Under just one iteration of the RG procedure in Eq. \ref{eq:RG}, $T_l$ goes into $\vec E_a^*$ and $T_r$ goes into $\vec E_b^*$.  
This means that, in the thermodynamic limit, $\epsilon_a(t)$ and $\epsilon_b(t)$ are both strictly zero at $t_c = \frac{\pi}{2}$ for the chain described by Eq. \ref{eq:J_i}. 
Thus, for $t$ in the vicinity of $t_c$, the maximum of $\abs{\epsilon_a(t)}$ and $\abs{\epsilon_b(t)}$ equals to $c \tau$ where $\tau = \abs{t - t_c}$ and $c$ is a nonzero constant.  
This then results in a logarithmic divergence in the rate function:  
\begin{equation}
  l_s(t) \propto -\log(\abs{t - t_c}), \hspace{5mm} \text{   for $t$ close to $t_c$} 
\end{equation}
The rate function of the chain described by Eq. \ref{eq:J_i} with $p = \frac{1}{2}$ is shown in Fig. \ref{fig:rate_log}.      
When $J_i$s are not exactly fine-tuned to be $1$ and $\frac{1}{2}$, the logarithmic divergence degrades into two highly-peaked linear-cusps around $t_c$. 
An example of this is also shown in Fig. \ref{fig:rate_log}, with $J_i = 1$ and $0.499$, each with probability $\frac{1}{2}$.     

We now generalize the result to other $N$s. 
First note that the commutativity of the Potts transfer matrices still holds for $N > 2$. 
Secondly, at least for $N = 3, 4, 5$ which has been explicitly studied in \cite{potts_rg}, multiplying the non-critical RG fixed-point transfer matrices with one another give the zero matrix.  
For example, when $N = 5$, there are three non-critical RG fixed-points, $\vec E_a^* = (1,1,1,1,1)$, $\vec E_b^* = (1, \frac{1}{4}(-1 + \sqrt{5}), \frac{1}{4}(-1-\sqrt{5}), \frac{1}{4}(-1-\sqrt{5}), \frac{1}{4}(-1+\sqrt{5}))$, and $\vec E_c^* = (1, \frac{1}{4}(-1-\sqrt{5}), \frac{1}{4}(-1+\sqrt{5}), \frac{1}{4}(-1+\sqrt{5}), \frac{1}{4}(-1-\sqrt{5}))$, corresponding to three fixed-point transfer matrices, $T_a^*, T_b^*, T_c^*$.
As one can check, $T_a^* T_b^* = T_b^* T_c^* = T_c^* T_a^* = 0$. 
Then, the arguments from Eq. \ref{eq:Ta} to Eq. \ref{eq:la} follow identically, giving 
\begin{equation}
  \qbraket{l_s(t)} = \min(l_a(t), l_b(t), l_c(t)) 
\end{equation}
where $l_a(t), l_b(t), l_c(t)$ are analogously defined as in Eq. \ref{eq:la}. 

In this paper, we discussed the effect of disorder on the DQPT of the Potts chain under the extreme dynamical quantum quench.   
The DQPT in the presence of disorder remains sharp in general, and is due to the competition between the non-critical phases of the pure model.       
This highlights the significance of these non-critical RG fixed points, which may seem inapparent in the pure model. 
For equilibrium phase transitions of the disordered systems, the Harris criterion \cite{harris} is a general rule that characterizes the effect of small disorder. 
It argues that if the fixed-point coupling distribution under the RG flow is the pure critical fixed-point, the critical exponents of the disordered system should be the same as those of the pure model.  
As seen, the RG picture of the DQPT here does not involve the critical phase of the pure model at all.  
In fact, in the pure model, the singularity in the rate function for some $N$s is not even controlled by a critical RG fixed point \cite{potts_rg}.   
The DQPT of the disordered Potts chain is thus incompatible with the Harris criterion, and a general rule on how DQPTs behave with disorder awaits. 

\begin{acknowledgments}
The author is grateful to Ling Wang for hosting him at the Beijing Computational Science Research Center, introducing him to DQPTs, and many stimulating discussions.  
He is also grateful for mentorship from his advisor Roberto Car at Princeton. 
The author acknowledges support from the DOE Award DE-SC0017865. 
\end{acknowledgments}
\bibliographystyle{apsrev}

\begin{thebibliography}{10}
\expandafter\ifx\csname natexlab\endcsname\relax\def\natexlab#1{#1}\fi
\expandafter\ifx\csname bibnamefont\endcsname\relax
  \def\bibnamefont#1{#1}\fi
\expandafter\ifx\csname bibfnamefont\endcsname\relax
  \def\bibfnamefont#1{#1}\fi
\expandafter\ifx\csname citenamefont\endcsname\relax
  \def\citenamefont#1{#1}\fi
\expandafter\ifx\csname url\endcsname\relax
  \def\url#1{\texttt{#1}}\fi
\expandafter\ifx\csname urlprefix\endcsname\relax\def\urlprefix{URL }\fi
\providecommand{\bibinfo}[2]{#2}
\providecommand{\eprint}[2][]{\url{#2}}

\bibitem[{\citenamefont{Heyl}(2018)}]{Heyl_2018}
\bibinfo{author}{\bibfnamefont{M.}~\bibnamefont{Heyl}},
  \bibinfo{journal}{Reports on Progress in Physics}
  \textbf{\bibinfo{volume}{81}}, \bibinfo{pages}{054001}
  (\bibinfo{year}{2018}).

\bibitem[{\citenamefont{Heyl et~al.}(2013)\citenamefont{Heyl, Polkovnikov, and
  Kehrein}}]{dqpt}
\bibinfo{author}{\bibfnamefont{M.}~\bibnamefont{Heyl}},
  \bibinfo{author}{\bibfnamefont{A.}~\bibnamefont{Polkovnikov}},
  \bibnamefont{and} \bibinfo{author}{\bibfnamefont{S.}~\bibnamefont{Kehrein}},
  \bibinfo{journal}{Phys. Rev. Lett.} \textbf{\bibinfo{volume}{110}},
  \bibinfo{pages}{135704} (\bibinfo{year}{2013}).

\bibitem[{\citenamefont{Zvyagin}(2016)}]{zvyagin}
\bibinfo{author}{\bibfnamefont{A.~A.} \bibnamefont{Zvyagin}},
  \bibinfo{journal}{Low Temperature Physics} \textbf{\bibinfo{volume}{42}},
  \bibinfo{pages}{971} (\bibinfo{year}{2016}).

\bibitem[{\citenamefont{{Cardy}}(1996)}]{cardy}
\bibinfo{author}{\bibfnamefont{J.}~\bibnamefont{{Cardy}}},
  \emph{\bibinfo{title}{{Scaling and Renormalization in Statistical Physics}}}
  (\bibinfo{year}{1996}).

\bibitem[{\citenamefont{{Wu}}(2019)}]{potts_rg}
\bibinfo{author}{\bibfnamefont{Y.}~\bibnamefont{{Wu}}}, \bibinfo{journal}{arXiv
  e-prints} \bibinfo{eid}{arXiv:1906.07945} (\bibinfo{year}{2019}),
  \eprint{1906.07945}.

\bibitem[{\citenamefont{Heyl}(2015)}]{scaling_tfic}
\bibinfo{author}{\bibfnamefont{M.}~\bibnamefont{Heyl}}, \bibinfo{journal}{Phys.
  Rev. Lett.} \textbf{\bibinfo{volume}{115}}, \bibinfo{pages}{140602}
  (\bibinfo{year}{2015}).

\bibitem[{\citenamefont{Karrasch and Schuricht}(2017)}]{dqpt_potts}
\bibinfo{author}{\bibfnamefont{C.}~\bibnamefont{Karrasch}} \bibnamefont{and}
  \bibinfo{author}{\bibfnamefont{D.}~\bibnamefont{Schuricht}},
  \bibinfo{journal}{Phys. Rev. B} \textbf{\bibinfo{volume}{95}},
  \bibinfo{pages}{075143} (\bibinfo{year}{2017}).

\bibitem[{\citenamefont{Harris and Lubensky}(1974)}]{lubensky}
\bibinfo{author}{\bibfnamefont{A.~B.} \bibnamefont{Harris}} \bibnamefont{and}
  \bibinfo{author}{\bibfnamefont{T.~C.} \bibnamefont{Lubensky}},
  \bibinfo{journal}{Phys. Rev. Lett.} \textbf{\bibinfo{volume}{33}},
  \bibinfo{pages}{1540} (\bibinfo{year}{1974}).

\bibitem[{\citenamefont{Sowa and Kaiser}()}]{ttmath}
\bibinfo{author}{\bibfnamefont{T.}~\bibnamefont{Sowa}} \bibnamefont{and}
  \bibinfo{author}{\bibfnamefont{C.}~\bibnamefont{Kaiser}},
  \bibinfo{note}{www.ttmath.org}.

\bibitem[{\citenamefont{Harris}(1974)}]{harris}
\bibinfo{author}{\bibfnamefont{A.~B.} \bibnamefont{Harris}},
  \bibinfo{journal}{Journal of Physics C: Solid State Physics}
  \textbf{\bibinfo{volume}{7}}, \bibinfo{pages}{1671} (\bibinfo{year}{1974}).

\end{thebibliography}

\end{document}